# Effective Schrödinger equation with general ordering ambiguity position-dependent mass Morse potential

**Sameer M. Ikhdair**
Physics Department, Near East University, Nicosia, North Cyprus, Turkey
*sikhdair@neu.edu.tr*
*Tel: +90-392-2236624; Fax: +90-392-2236622*

**Abstract**

We solve the parametric generalized effective Schrödinger equation with a specific choice of position-dependent mass function and Morse oscillator potential by means of the Nikiforov-Uvarov (NU) method combined with the Pekeris approximation scheme. All bound-state energies are found explicitly and all corresponding radial wave functions are built analytically. We choose the Weyl or Li and Kuhn ordering for the ambiguity parameters in our numerical work to calculate the energy spectrum for a few $H_2$, $LiH$, $HCl$ and $CO$ diatomic molecules with arbitrary vibration $n$ and rotation $l$ quantum numbers and different position-dependent mass functions. Two special cases including the constant mass and the vibration s-wave ($l=0$) are also investigated.

## 1. Introduction

The Morse oscillator potential [1] plays an important physical role in quantum mechanics and in the field of molecular physics describing the vibrations between two atoms and has attracted a great interest for decades. It is an exactly solvable potential for s-waves ($l=0$) and is of much use in spectroscopic applications. However, for the rotating Morse potential ($l \neq 0$), some semiclassical and/or numerical solutions have been obtained by using various approximation methods. An effective approximation commonly known as the Pekeris approximation [2] was used to obtain the semiclassical solutions [3-12]. This approximation was employed for solving the rotating Morse potential with any $l$-states [7,10]. These methods include the variational [3], supersymmetry (SUSY) [4], the hypervirial perturbation [5], the shifted 1/N expansion (SE) and the modified shifted 1/N expansion (MSE) [6], the Nikiforov-Uvarov (NU) method [7], the asymptotic iteration method (AIM) [9], the exact quantization rule (EQR) [10], the tridiagonal J--matrix representation (TJM) [11], the Wigner distribution function (WDF) for the rotating Morse oscillator (RMO) [12], the two-point quasi-rational approximant technique (TQA) [13] and the results of Duff and Rabitz (DR) [14] etc.

Over the past few years, the effects of the position-dependent mass (PDM) on the solutions of the non-relativistic and relativistic wave equations have been of considerable current interest [15-21]. Indeed, there exists a wide variety of situations in which PDM is of utmost relevance. PDM also holds out to deformations in the quantum canonical commutation relations or curvature of the underlying space. This subject has wide applications in the study of the material science and condensed matter physics, such as quantum semiconductors [22], the electronic properties of quantum wells and quantum dots (QDs) [23], $^3He$ clusters [24], quantum liquids [25], graded alloys and semiconductor heterostructures [26] and the impurities in crystals [27] etc.

The point canonical transformation [15], deformed algebras [28], the quadratic algebra method [29], the path-integral method [30], Lie algebra approach [31] and SUSY formalism [32] are a few methods used to solve exactly the wave equations for the case of constant and PDM distributions.

In the last decade, the algebraic techniques as the Lie algebraic methods [33] and factorization methods [34] have been developed and applied in molecular spectroscopy. The Lie algebraic methods have been introduced in the systematic study of spectra of molecules (vibron model) [35]. The introduction was based on the second quantization of the Schrödinger equation with a three dimensional Morse potential and described rotation-vibration spectra of diatomic molecules [36]. Soon afterwards the algebraic method was extended to rotation-vibration spectra of polyatomic molecules [37]. The Morse oscillator can be solved exactly using a variety of algebraic methods [34]. These problems correspond to different realizations of the so(2,1) algebra and a comparison of generators of the algebra may be used to identify mappings between each pair of systems. The resultant transition operators act as ladder, or energy changing, operators in the cases of the Coulomb and harmonic oscillator potentials, whereas they act as shift operator, acting at constant energy, in the case of the Morse potential [38]. A realization of the raising and lowering operators for the Morse potential has also been given. They satisfy the commutation relations for the SU(2) group [38]. An alternative algebraic approach, the use of the method of supersymmetric quantum mechanics, or factorization, produces in each case a set of shift operators. The bound-state solutions and the su(1,1) description of the $D$-dimensional radial harmonic oscillator, the Morse and the $D$-dimensional radial Coulomb Schrödinger equations are reviewed in a unified way using the point canonical transformation method, it is established that the spectrum generating su(1,1) algebra for the first problem is converted into a potential algebra for remaining two. This analysis is also extended to Schrödinger equations containing some position-dependent mass [39]. The algebraic solutions of the Schrödinger equation with position-dependent mass for the Morse potential are obtained by the series expansion method. The Morse potential and the PDM, $m(r) = m_0 e^{\lambda r}$, are expanded in the series about the origin [40]. A singular oscillator Hamiltonian with a position-dependent effective mass has been constructed. It was found out that an su(1,1) algebra is the hidden symmetry of the quantum system and the isospectral potentials $V(x)$ depend on the different choices of the $m(x)$ [40].

Recently, the Nikiforov-Uvarov (NU) method [41] and other methods were used to solve the Schrödinger [16-19], the Klein-Gordon (KG) [42,43] and the Dirac [44] equations with PDM distribution and constant mass [45-47].

Quite recently, we solved the PDM Schrödinger equation [48] with a suitable choice of PDM function very similar to the ansatze used by Bagchi *et al.* in Eq. (10) of Ref. [49] which worked well for the Morse oscillator potential ($l \neq 0$). Further, the energy spectrum and their corresponding wave functions are obtained for the present system. In our solution, we applied parametric generalization model of the NU method [44,46] combined with Pekeris approximation scheme for the centrifugal term $l(l+1)/r^2$ [48].

To the best of our knowledge, the rotation-vibration motion of the diatomic molecules have not been studied in the coordinate-dependent ordering Schrödinger equation with general parameters $\alpha, \beta, \gamma$ and $a$ satisfying the condition $\alpha + \beta + \gamma = -1$. For the sake of completeness, the aim of the present work is to extend the NU solution of Ref. [48] to a general ordering ambiguity PDM radial Schrödinger equation [50-54] with the effective potential $V_{\text{eff.}}(r) = V_M(r) + U_{\alpha\beta\gamma a}(r)$, where $V_M(r)$ is the Morse oscillator potential and $U_{\alpha\beta\gamma a}(r)$ is the ordering potential. The present solution comprises all the well-known ordering of parameters proposed by many authors in the literature like the ordering of Gora and Williams [52] ($a + \gamma = 0, \alpha = -1$), BenDaniel and Duke [51] ($a = \alpha = \gamma = 0$), Zhu and Kroemer [53] ($a = 0, \alpha = -1/2 = \gamma$), Li and Kuhn [54] ($a = 0, \alpha = 0, \gamma = -1/2$) and Weyl [50] ($a = 1, \alpha = 0 = \gamma$). However, the solution introduced in [48] is valid only for the Weyl ordering [50]. It is worthy to note that the work in Ref. [48] introduced the present approach for a constant mass and is now applied to a non-constant mass case. However, the present analytical solution is presented in terms of the general parameters $\alpha, \beta, \gamma$ and $a$ satisfying the condition $\alpha + \beta + \gamma = -1$ and computed numerically for the Weyl ordering [50].

The contents of the paper are as follows. In Sect. 2, we obtain the NU approximate analytic bound state solution of the general coordinate-dependent PDM Schrödinger equation with the Morse oscillator potential using the Pekeris approximation scheme. Two special cases of much interest like the constant mass and the vibration ($l = 0$) are investigated. In Sect. 3, we compute the rotation-vibration bound-state energies for $H_2$, $LiH$, $HCl$ and $CO$ molecules for different vibration $n$ and rotation $l$ quantum numbers considering the constant and varying mass cases. Conclusions are presented in Sect. 4.

## 2. Morse Oscillator Potential with Position-Dependent Mass

Now in our analytic implementation we use the classical Morse oscillator potential model for diatomic molecules defined by [1]:

$$V(r) = V_1 e^{-2b(r-r_e)} - V_2 e^{-b(r-r_e)}, \tag{1}$$

where $V_1 > 0$ and $V_2 > 0$ are two strength parameters of the potential corresponding to $D_e$ and $2D_e$, respectively. The potential strength $D_e$ is the dissociation energy, the parameters $r_e$ and $b$ are two positive parameters to signify the equilibrium position of the nuclei and the width of the potential well, respectively. The numerical values of these parameters are shown in Table 1 for different diatomic molecules along with the sources from which these data were extracted. The vibrations of a two-atomic molecule can be excellently described by this potential type for s-waves ($l = 0$). If we want to calculate the rotation bound state energy for $l \neq 0$, we need to make approximation to the centrifugal term. We begin by defining a quite general Hermitian effective Hamiltonian for the case of a spatially dependent mass. We consider the general parameterized Schrödinger Hamiltonian being expressed in terms of the general coordinate dependent ordering of parameters $\alpha, \beta, \gamma$ and $a$ satisfying the condition $\alpha + \beta + \gamma = -1$. For example, the specific choice of the parameters $(a = 1, \alpha = 0 = \gamma)$ leads us to the Weyl ordering [48]. This parameterized effective Hamiltonian with four terms was proposed by von Roos [55] including the case of the Weyl ordering [50] and is given by [56]

$$H = \frac{1}{4(a+1)} \left\langle a\left[m^{-1}(\vec{r})\hat{p}^2 + \hat{p}^2 m^{-1}(\vec{r})\right] + m^{\alpha}(\vec{r})\hat{p}m^{\beta}(\vec{r})\hat{p}m^{\gamma}(\vec{r}) + m^{\gamma}(\vec{r})\hat{p}m^{\beta}(\vec{r})\hat{p}m^{\alpha}(\vec{r}) \right\rangle$$

where $\alpha, \beta, \gamma$ and $a$ are the called ambiguity parameters satisfying the constraint $\alpha + \beta + \gamma = -1$. The above parameterized Hamiltonian is valid for all mentioned types of ambiguity orderings. A similar Hamiltonian was used by Levinger and collaborators [57]. Using the properties of the canonical commutators, it is easy to show that one can put the momenta to the right, so obtaining the following effective Hamiltonian

$$H = \frac{1}{2m}\hat{p}^2 + \frac{i\hbar}{2}\frac{m'}{m^2}\hat{p} + U_{\alpha\beta\gamma a}(r),$$

$$U_{\alpha\beta\gamma a}(r) = -\frac{\hbar^2}{4m^3(a+1)}\left[(\alpha + \gamma - a)mm'' + 2(a - \alpha - \gamma - \alpha\gamma)m'^2\right], \tag{2}$$

where $m = m(r)$ is real mass function and prime denotes the differentiation with respect to $r$. It is worth to mention that all the ambiguity is in the last effective potential term can be eliminated by imposing some convenient constraints over the ambiguity parameters: $\alpha + \gamma - a = 0$, $a - \alpha - \gamma - \alpha\gamma = 0$ which have two possible solutions (i) $\alpha = 0$ and $a = \gamma$ or (ii) $a = \alpha$ and $\gamma = 0$. In this case the effective Schrödinger equation will not depend on the ambiguity parameters but will contain a first order derivative term. From the above effective Hamiltonian, we write down the Schrödinger equation for any radial potential function as [56]

$$-\frac{\hbar^2}{2m}\left[\left(\frac{d^2\psi(r)}{dr^2} - \frac{l(l+1)}{r^2}\right) - \frac{m'}{m}\frac{d\psi(r)}{dr}\right] + [V(r) + U_{\alpha\beta\gamma a}(r) - E]\psi(r) = 0. \tag{3}$$

To eliminate the first derivative in Eq. (3), we shall use a simple method based on the point canonical transformations (PCT) technique which maps the ordinary Schrödinger equation with a constant mass (reference problem) to a Schrödinger equation with a spatially dependent mass (target). This procedure has been used recently to obtain solutions for particular potentials [15,58]. For that, substituting

$$\psi(r) = \sqrt{m}\varphi(r), \tag{4}$$

into Eq. (3), we obtain, after some algebra, a differential equation in a more familiar form

$$-\frac{\hbar^2}{2m}\frac{d^2\varphi(r)}{dr^2} + (U_{\text{eff}}(r) - E)\varphi(r) = 0, \tag{5}$$

with the effective potential

$$U_{\text{eff}}(r) = V_1 e^{-2b(r-r_e)} - V_2 e^{-b(r-r_e)} + U_{\alpha\beta\gamma a}(r) + \frac{\hbar^2}{4m}\left[\frac{3}{2}\left(\frac{m'}{m}\right) - \frac{m''}{m}\right] + \frac{l(l+1)\hbar^2}{2mr^2}, \tag{6}$$

where $r \in (0,\infty)$ and $\varphi(0) = \varphi(\infty) = 0$. It is found that the Weyl ambiguity ordering [50] is equivalent to Li and Kuhn ambiguity ordering [54] in the case of the Schrödinger equation.

Let us now come to the analytical solution of Eq. (5) together with Eq. (6) where the mass is allowed to depend on the position $r$ [15,28,48,58,59]. Such systems are motivated mostly by condensed matter problems [60] and mathematical physics ones, such as the quest of solutions for Schrödinger equation [28] and scattering in abrupt hetero structures. For instance, Schmidt [61] studied the quantum evolution and temporal evolution of wave packets revival which have variable mass and are confined in an infinite potential well. This effect takes place also in other systems, as an example, as the circular setup presented in [62], and the 2D billiard in the presence of a magnetic field [63]. For that, we consider the analytically solvable model [15] where the effective mass is taken in the form [48,49,64]

$$m(r) = m_0 \left(1 - \varepsilon e^{-b(r-r_e)}\right)^{-2}, \quad m(\infty) = m_0, \tag{7}$$

with $0 \leq \varepsilon < 1$ being a real coupling constant parameter. The above mass function $m(r) = m_0\left(1 - \varepsilon_0 e^{-br}\right)^{-2}$, $\varepsilon_0 = \varepsilon e^{br_e}$ is convergent $m(r) \to m_0$ when $r \to \infty$ and finite $m(r) = m_0(1-\varepsilon_0)^{-2}$ when $r \to 0$. It is found that the mass function has the exponential form of the reciprocal Morse-like potential as conjectured by the non-relativistic (Eq. (5) of Ref. [49] and Eq. (6) of Ref. [65]) and relativistic (Eq. (30) of [66]) models. This is because of the dominating potential field in the region between the two interacting nuclei beyond the fact that it results in a solvable wave equation (mathematical suitability). We thus find out that the Morse oscillator potential $V(r)$ with reciprocal Morse-like PDM (7) can be easily reduced to it's standard form with constant mass. However, in the present choice of the mass functions, there is no loss of generality when the value of parameter $\varepsilon$ is taken to be small (i.e., $\varepsilon \to 0$). It must be stressed that other choice of mass functions can be made which work well for the Morse potential. It is also interesting to observe that some authors have used different form for the mass function, i.e., $m(x) = m_0 + m_1 e^{-\beta x} + m_2 e^{-2\beta x}$, which is simply obtained from the solution of Dirac equation for the Morse potential and can be easily reduced to the constant mass when $m_1 = m_2 = 0$ and convergent to $m_0$ when $x \to \infty$ [66]. On the other hand, Schmidt [61] selected three particular cases for the mass function $m(x) = cx^\alpha$ with $c = 1$ and $\alpha = 1, 2$ and $4$ to study the temporal evolution of a free wave-pucket inside an infinite well potential where it's mass is position dependent.

We will consider the Morse oscillator potential with non-vanishing orbital angular momentum quantum numbers and apply the parametric generalization of the NU method presented in section 2 and Appendix A of Ref. [48] to calculate the energy spectrum and the corresponding wave functions. The Morse oscillator potential with $l \neq 0$ is not exactly solvable and hence our numerical results cannot be checked against exact ones. However, many numerical and perturbative results have been published in recent years [7,9,10,67]. The most widely used approximation was devised by Pekeris [2] which is based on the expansion of the centrifugal term $l(l+1)r^{-2}$ in a series of exponential terms around the equilibrium inter-nuclear position $r = r_e$ ($x = 0$) of the Morse oscillator potential by keeping terms up to second order $r/r_e$ (i.e., at low excitation energy, where $r \approx r_e$). Other approximations have also been devised but they require numerical solution of transcendental equations [6,68,69]. The Pekeris approximation is mainly based on an expansion in powers of exponential function and truncated at the quartic term. It is effective only in approximating the lower-excitation rotation energy states but quite poor in the description of higher-excitation rotation energy states because of the large interatomic separations [44]. To apply the Pekeris approximation, we change the parameters and coordinates: $v = br_e$, $x = (r - r_e)/r_e$ and $d/dr = (1/r_e)(d/dx)$ to obtain [7]:

$$V_{\rm rot}(x) = \frac{l(l+1)}{r_e^2}\left(D_0 + D_1 e^{-vx} + D_2 e^{-2vx}\right), \quad x \in (0, \infty) \tag{8}$$

where

$$D_0 = 1 - \frac{3}{v} + \frac{3}{v^2}, \quad D_1 = \frac{4}{v} - \frac{6}{v^2}, \quad D_2 = -\frac{1}{v} + \frac{3}{v^2}. \tag{9}$$

Also we can rewrite Eq. (5) more explicitly as

$$\frac{\left(1 - \varepsilon e^{-vx}\right)^2}{r_e^2}\frac{d^2\varphi(x)}{dx^2} - \frac{2m_0}{\hbar^2}\left(U_{\rm eff}(x) - E\right)\varphi(x) = 0, \tag{10}$$

where we have used $\varphi(x) = \varphi(r)$ and the identifications:

$$U_{\rm eff}(x) = -\frac{\hbar^2}{2m_0 r_e^2}\left(-a_0 + a_1 e^{-vx} - a_2 e^{-2vx}\right)$$

$$a_0 = l(l+1)D_0, \quad a_1 = -l(l+1)D_1 + \frac{2m_0 r_e^2}{\hbar^2}V_2 + v^2 r_e^2\left(\frac{\alpha+\gamma+1}{a+1}\right)\varepsilon,$$

$$a_2 = l(l+1)D_2 + \frac{2m_0 r_e^2}{\hbar^2}V_1 + v^2 r_e^2\left(\frac{1+2\alpha+2\gamma+4\alpha\gamma-a}{a+1}\right)\varepsilon^2. \tag{11}$$

Making further change of variables $z = e^{-vx} \in [0,1]$ followed by little straightforward algebra, we can finally reduce Eq. (10) with the aid of Eq. (11) to a hypergeometric-type equation:

$$\varphi''(z) + \frac{(1-\varepsilon z)}{z(1-\varepsilon z)}\varphi'(z) + \frac{1}{z^2(1-\varepsilon z)^2}\left(-\gamma_1 z^2 + \gamma_2 z - \varepsilon_{nl}^2\right)\varphi(z) = 0, \tag{12}$$

where $\varphi(z) = \varphi(x)$ and

$$\varepsilon_{nl}^2 = -\frac{2m_0 r_e^2}{\hbar^2 b^2}E + \frac{l(l+1)}{b^2 r_e^2}D_0, \tag{13a}$$

$$\gamma_1 = \frac{2m_0 r_e^2}{\hbar^2 \beta^2}V_1 + \frac{l(l+1)}{\beta^2}D_2 + \left(\frac{1+2\alpha+2\gamma+4\alpha\gamma-a}{a+1}\right)r_e^2 \varepsilon^2, \tag{13b}$$

$$\gamma_2 = \frac{2m_0 r_e^2}{\hbar^2 \beta^2}V_2 - \frac{l(l+1)}{\beta^2}D_1 + r_e^2\left(\frac{\alpha+\gamma+1}{a+1}\right)\varepsilon. \tag{13c}$$

Comparing Eq. (12) with Eq. (1) of Ref. [48] and applying the parametric generalization of the NU introduced in the Appendix A of Ref. [48], it gives the following values for the set of polynomials and constants as

$$\tilde{\tau}(z) = 1 - \varepsilon z, \quad \sigma(z) = z(1-\varepsilon z), \quad \tilde{\sigma}(z) = -\gamma_1 z^2 + \gamma_2 z - \varepsilon_{nl}^2 \tag{14}$$

$$c_1 = 1, \; c_2 = c_3 = \varepsilon, \; c_4 = 0, \; c_5 = -\frac{\varepsilon}{2}, \; c_6 = \gamma_1 + \frac{\varepsilon^2}{4}, \; c_7 = -\gamma_2, \; c_8 = \varepsilon_{nl}^2,$$

$$c_9 = \frac{\varepsilon^2}{4}\left[1 + 4\varepsilon_{nl}^2 + \frac{4}{\varepsilon}\left(\frac{\gamma_1}{\varepsilon} - \gamma_2\right)\right], \; c_{10} = 2\varepsilon_{nl}, \; c_{11} = S = \sqrt{1 + 4\varepsilon_{nl}^2 + \frac{4}{\varepsilon}\left(\frac{\gamma_1}{\varepsilon} - \gamma_2\right)},$$

$$c_{12} = \varepsilon_{nl}, \; c_{13} = \frac{1}{2}(1+S), \; A = \gamma_1, \; B = \gamma_2, \; C = \varepsilon_{nl}^2. \tag{15}$$

Further, the application of the relations (A1-A4) of Ref. [48] together with the above values of constants gives the following particular values for the essential functions:

$$\pi(z) = \varepsilon_{nl} - \frac{\varepsilon}{2}(1 + 2\varepsilon_{nl} + S)z, \tag{16}$$

$$k = \gamma_2 - 2\varepsilon \varepsilon_{nl}^2 - \varepsilon \varepsilon_{nl} S, \tag{17}$$

and

$$\tau(z) = 1 + 2\varepsilon_{nl} - \varepsilon(2 + 2\varepsilon_{nl} + S)z, \tag{18}$$

where $\tau'(z) = -\varepsilon(2 + 2\varepsilon_{nl} + S) < 0$. Also, applying the relation A5, the energy equation for the potential model (1) can be found as

$$\varepsilon_{nl} = \frac{n(n+1)\varepsilon - (2n+1)\sqrt{\gamma_1} + \gamma_2}{2\sqrt{\gamma_1} - (2n+1)\varepsilon}, \; 0 \le n \le n_{max}, \tag{19}$$

where $n$ is the vibration quantum number and $n_{max}$ denotes the maximum number of bound states where sign changes. The above equation can be alternatively expressed as

$$E_{nl} = B_0\left(1 - \frac{3}{br_e} + \frac{3}{b^2 r_e^2}\right) - A_0\left[Q_2 - \sqrt{A_0}\left(n + \frac{1}{2}\right)\varepsilon\right]^{-2}$$

$$\times \left\{\frac{1}{2}\sqrt{A_0}\left[\left(n + \frac{1}{2}\right)^2 - \frac{1}{4}\right]\varepsilon + \sqrt{\frac{1}{A_0}}Q_1 - \left(n + \frac{1}{2}\right)Q_2\right\}^2, \qquad (20)$$

where $A_0 = \hbar^2 b^2 /(2m_0)$, and $B_0 = \hbar^2 l(l+1)/(2m_0 r_e^2)$. Further, $Q_1$ and $Q_2$ are expressed in terms of the general ordering parameters $\alpha, \gamma$ and $a$ together with the mass coupling constant $\varepsilon$ as

$$Q_1 = D_e - B_0\left(\frac{2}{br_e} - \frac{3}{b^2 r_e^2}\right) + \frac{r_e^2 A_0}{2}\left(\frac{\alpha + \gamma + 1}{a + 1}\right)\varepsilon, \qquad (21a)$$

$$Q_2 = \sqrt{D_e - B_0\left(\frac{1}{br_e} - \frac{3}{b^2 r_e^2}\right) + r_e^2 A_0\left(\frac{1 + 2\alpha + 2\gamma + 4\alpha\gamma - a}{a + 1}\right)\varepsilon^2}. \qquad (21b)$$

where $l = 0, 1, 2, \cdots$, signify the rotation quantum number. Indeed, the energy expression in (20) is the most general form valid for any ordering of parameters. It is worth to note that the results of Ref. [48] can be easily recovered when the values of ordering parameters take the values $a = 1$ and $\alpha = 0 = \gamma$ (Weyl ordering) [50]. For the s-waves ($l = 0$), the exact vibration energy states turn to be

$$E_{nl} = -A_0\left[q_2 - \sqrt{A_0}\left(n + \frac{1}{2}\right)\varepsilon\right]^{-2}\left\{\frac{1}{2}\sqrt{A_0}\left[\left(n + \frac{1}{2}\right)^2 - \frac{1}{4}\right]\varepsilon + \sqrt{\frac{1}{A_0}}q_1 - \left(n + \frac{1}{2}\right)q_2\right\}^2, \qquad (22)$$

where $n = 0, 1, 2, \cdots n_{max}$ and the parameters $q_1$ and $q_2$ are being expressed in terms of the general ordering parameters as

$$q_1 = D_e + \frac{r_e^2 A_0}{2}\left(\frac{\alpha + \gamma + 1}{a + 1}\right)\varepsilon, \qquad (23a)$$

$$q_2 = \sqrt{D_e + r_e^2 A_0\left(\frac{1 + 2\alpha + 2\gamma + 4\alpha\gamma - a}{a + 1}\right)\varepsilon^2}. \qquad (23b)$$

The above energy formula represents a general solution containing the mass coupling parameter $\varepsilon$ which is shaping the mass function (7). For example, when the mass is constant, i.e., $\varepsilon = 0$, then the energy spectrum turns to become

$$\varepsilon_{nl} = \frac{1}{2}\frac{\gamma_2}{\sqrt{\gamma_1}} - \left(n + \frac{1}{2}\right), \qquad (24)$$

or it can be explicitly expressed as

$$E_{nl} = B_0\left(1 - \frac{3}{\beta} + \frac{3}{\beta^2}\right) - A_0\left[\frac{p_1}{\sqrt{p_2}} - \left(n + \frac{1}{2}\right)\right]^2, n = 0, 1, 2, \cdots, n_{max}, \; l = 0, 1, 2, \cdots$$

$$p_1 = \frac{D_e}{A_0} - \frac{l(l+1)}{\beta^2}\left(\frac{2}{\beta} - \frac{3}{\beta^2}\right) \text{ and } p_2 = \frac{D_e}{A_0} - \frac{l(l+1)}{\beta^2}\left(\frac{1}{\beta} - \frac{3}{\beta^2}\right). \qquad (25)$$

which is the rotation-vibration energy spectrum for the Morse oscillator potential. Consequently, the vibration energy spectrum for s-wave ($l = 0$) is

$$E_n = -A_0\left[\sqrt{\frac{D_e}{A_0}} - \left(n + \frac{1}{2}\right)\right]^2, n = 0, 1, 2, \cdots, n_{max} \leq \frac{1}{2}\left(\sqrt{\frac{4D_e}{A_0}} - 1\right), \qquad (26)$$

where $n_{max}$ is the integer number of bound states for the whole bound spectrum near the continuous zone. The above energy spectrum coincides with those ones given in literature [11].

In what follows we turn to the calculation of the corresponding wave functions. The wave functions can now be obtained from the relations (A6-A10) in Appendix A of Ref. [48]. At first, the weight function can be found as [18,41]

$$\rho(z) = z^{2\varepsilon_{nl}}(1-\varepsilon z)^S,$$

where $\varepsilon_{nl}$ is given in Eq. (19) which is in turn gives the first part of the wave functions

$$y_n(z) \to P_n^{(2\varepsilon_{nl}, S)}(1-2\varepsilon z), \quad (27)$$

and the second part as

$$\phi(z) = z^{\varepsilon_{nl}}(1-\varepsilon z)^{(1+S)/2}. \quad (28)$$

Combining Eqs. (27) and (28), we obtain the unnormalized wave function being expressed in terms of the Jacobi polynomials as

$$\varphi(z) = N_{nl} \frac{n!\Gamma(2\varepsilon_{nl}+1)}{\Gamma(n+2\varepsilon_{nl}+1)} z^{\varepsilon_{nl}}(1-\varepsilon z)^{(1+S)/2} P_n^{(2\varepsilon_{nl}, S)}(1-2\varepsilon z). \quad (29)$$

On the other hand, we give the relation linking the hypergeometric function and the Jacobi polynomials (see formula 8.962.1) in [70]

$$_2F_1\left(-n, n+\nu+\mu+1, \nu+1; \frac{1-x}{2}\right) = \frac{n!\Gamma(\nu+1)}{\Gamma(n+\nu+1)} P_n^{(\nu, \mu)}(x),$$

to rewrite the radial wave functions as

$$\phi(r) = N_{nl} \frac{n!\Gamma(2\varepsilon_{nl}+1)}{\Gamma(n+2\varepsilon_{nl}+1)} e^{-b\varepsilon_{nl}(r-r_e)}(1-\varepsilon e^{-b(r-r_e)})^{(S+1)/2} {}_2F_1\left(-n, n+2\varepsilon_{nl}+S+1, 2\varepsilon_{nl}+1; \varepsilon e^{-b(r-r_e)}\right), \quad (30)$$

and the total normalized radial wave functions of the ordinary Morse oscillator potential are

$$\psi_{nl}(r) = N_{nl} \frac{n!\Gamma(2\varepsilon_{nl}+1)}{\Gamma(n+2\varepsilon_{nl}+1)} e^{-b\varepsilon_{nl}(r-r_e)}(1-\varepsilon e^{-b(r-r_e)})^{(S-1)/2} {}_2F_1\left(-n, n+2\varepsilon_{nl}+S+1, 2\varepsilon_{nl}+1; \varepsilon e^{-b(r-r_e)}\right), \quad (31)$$

where $0 \le \varepsilon < 1$ and the normalization constant

$$N_{nl} = \frac{1}{\Gamma(2\varepsilon_{nl}+1)} \left[\frac{n\Gamma(S+2)}{2b\Gamma(n+2\varepsilon_{nl}+1)} \sum_{m=0}^{\infty} \frac{(-1)^m (n+2\varepsilon_{nl}+S+1)_m \Gamma(n+m)}{m!(m+2\varepsilon_{nl})!\Gamma(m+2\varepsilon_{nl}+S+2)} f_{nl}\right]^{-1/2},$$

and

$$f_{nl} = {}_3F_2(2\varepsilon_{nl}+m, -n, n+2\varepsilon_{nl}+S+1, 2\varepsilon_{nl}+m+S+2;1),$$

where $S = c_{11}$ is defined in Eq. (15). It should be noted that the above solutions are well-behaved at the boundaries, i.e., a regular solution near the origin could be $\psi_{nl}(r \to 0) \to e^{b\varepsilon_{nl}r_e}(1-\varepsilon e^{br_e})^{(S-1)/2}$ and asymptotically at infinity as $\psi_{nl}(r \to \infty) \to 0$.

When the mass becomes constant, we treat this case separately as follows. The weight function given by Eq. (5) of Ref. [48] can be found as

$$\rho(z) = z^{2\varepsilon_{nl}} e^{-2\sqrt{\gamma_1}z}, \quad (32)$$

which leads to the Laguerre polynomials:

$$y_n(z) \to z^{-2\varepsilon_{nl}} e^{2\sqrt{\gamma_1}z} \frac{d^n}{dz^n}\left(z^{n+2\varepsilon_{nl}} e^{-2\sqrt{\gamma_1}z}\right) \to L_n^{2\varepsilon_{nl}}(x), \quad (33)$$

where $x = 2\sqrt{\gamma_1}z$. The second part of the wave functions given by Eq. (7) of Ref. [48] can be found as

$$\phi(z) = z^{\varepsilon_{nl}} e^{-\sqrt{\gamma_1} z}. \tag{34}$$

Hence, the unnormalized wave functions expressed in terms of the Jacobi polynomials read as

$$\psi_{nl}(r) = B_{nl} \frac{1}{r} \left(2\sqrt{\gamma_1}\right)^{-\varepsilon_{nl}} x^{\varepsilon_{nl}} e^{-x/2} L_n^{2\varepsilon_{nl}}(x), \tag{35}$$

where $x = 2\sqrt{\gamma_1} e^{-b(r-r_e)}$ and $B_{nl}$ is the normalization constant [7].

## 3. Results and Discussions

The Morse oscillator potential for $H_2$ and $LiH$ are plotted in Figure 1. Obviously, at the long interaction range $0 \leq r \leq 1.2 A°$, the potential interaction for the $LiH$ molecule is higher than that of $H_2$ molecule (i.e., long range weak interaction and positive bound state energy levels). At $r$ close to zero, it is nearly $60\,eV$ and $40\,eV$ for $LiH$ and $H_2$ molecules, respectively. However, in the interaction range $r \geq 1.4\,A°$, the potential interaction seems to be constant for the two molecules and approaches zero but are higher for $H_2$ molecule. Similar plot of the Morse potential for $HCl$ and $CO$ is shown in Figure 2. It is worth observing that at short distances in the range of $0 \leq r \leq 0.6 A°$, the $CO$ molecule has a stronger interaction than that of the $HCl$ molecule (i.e., short range strong interaction and positive bound state energy levels). Hence, at $r$ close to zero, it is nearly $1600\,eV$ and $400\,eV$ for $CO$ and $HCl$ molecules, respectively. However, in the interaction range $r \geq 0.6\,A°$, the two potential interactions corresponding the two molecules are found to coincide and constant. The numerically generated vibration s-bound state energies for $H_2$, $LiH$, $HCl$ and $CO$ molecules for the whole spectrum together with the number of bound states are shown in Table 2. These numerical computations were performed using the model parameters shown in Table 1 with the order of the eigenvalues represented by $n$ (the vibration quantum number). We have also seen that these vibration energy states are in excellent agreement with the exact results computed from analytic formula (26) for low- and high-level excitations. In Table 2, we had to go to higher-order bound states to reach a level at which our numerical results start deviating from exact results. These results are also compared with those ones obtained before by tridiagonal J-matrix representation [11]. The high agreement between the numerical and exact results is seen at low vibrations. We have also calculated numerically the number of bound states $n_{\max}$ along with the whole vibration bound-state spectrum near the continuous zone for each molecule. They are identical with numerical $n_{\max}$ values calculated in Table 2. In Figure 3, we plot the s-wave ($l = 0$) energy curves versus vibration quantum number $n$ for $H_2$, $HCl$ and $LiH$ molecules, respectively. Similar plot for the s-wave energy spectrum is shown for the $CO$ molecule in Figure 4. It should be noted that the calculated values of $n_{\max}$ obtained from Eq. (26) are as follows: 16, 24, 28 and 82 for $H_2$, $HCl$, $LiH$ and $CO$ molecules, respectively (see Figure 3 and Figure 4). The agreement between our numerical results and those generated by other methods up to four significant digits is reassuring. A slight difference could be seen and this is simply due to the Pekeris approximation [2] where the centrifugal term is being approximated to second order in $r/r_e$ (i.e., at low excitation energy, where $r \approx r_e$). If the average distance between the two nuclei in a diatomic molecule is $r_e$ then this can also be considered as being the order of magnitude of the uncertainty in the position of the valence electrons that are responsible for the binding of the nuclei. The Weyl ambiguity ordering choice demands $a = 1$ and $\alpha = 0 = \gamma$ in Eqs. (20) and (21), leading to same results obtained previously in Ref. [48] for $\varepsilon = 0$ (i.e., constant mass case, $m = m_0$). The extension of our study, choosing the Weyl ambiguity ordering, to the numerical calculations of the energy levels taking into account the position-dependent mass (PDM) with an arbitrary choice $\varepsilon \neq 0$ enables one to notice the variation of the energy values between the varying and constant mass cases. As an illustration, we display in Tables 3, 4, 5 and 6 for various values of the coupling parameter $\varepsilon$, the variation of the energy levels of the $H_2$, $LiH$, $HCl$ and $CO$ molecules, respectively using the model parameters in Table 1. The energy levels increase with the increase of $\varepsilon$ as demonstrated in Eq. (7). The numerical energy states corresponding for varying mass case and presented in Tables 3, 4, 5 and 6 can be tested by means of a mathematical software programs (Mathematica, Maple, Mathcad, ...etc) for a given set of parameters. A first look at the results shows that the bound state energy spectroscopy is the lowest (largest) for the $LiH$ ($CO$) molecule. However, the spectroscopes of $H_2$ and $HCl$ molecules are found to be intermediate and very close to one another. The free wave-packet inside the Morse oscillator potential where it's mass is position dependent [61] in turn leads to a respective change in the energy spectroscopy for every molecule.

To the best of our knowledge, this investigation represents the first explore to the interaction in the particles with a choice of specific mass functions moving in specified physical potentials. Since in many physical situations, particular mass distribution may be approximated by some typical functions like Eq. (7) and others mentioned in the present work, our results may provide the basic solutions for more complicated potential functions and PDM distributions.

## 4. Summary and Conclusions

To summarize, we have applied the parametric generalization of the NU method derived for the exponential-type potentials to obtain the bound state solutions of the effective Schrödinger equation with position dependent mass for the ordinary Morse oscillator potential for $l > 0$. The main feature of the present study is that our solution to the coordinate-dependent ordering Schrödinger equation with general parameters $\alpha, \beta, \gamma$ and $a$ satisfying the condition $\alpha + \beta + \gamma = -1$ can be easily reduced to the PDM Schrödinger equation in Ref. [48] once the Weyl ambiguity ordering $a = 1$ and $\alpha = 0 = \gamma$ is being applied in our numerical study. Furthermore, a suitable choice of a position mass function of the Morse-like form has also been devised. The present calculations include energy eigenvalues and the normalized wave functions expressed in terms of the Jacobi polynomials. This is a new feature due to the PDM environment since it can be seen from Eq. (7) that there are some restrictions exist on the parameter $\varepsilon$ in the effective mass function. In addition, since the energy eigenvalues of the realistic diatomic potentials are more accurately modified by the mass function, we are confident that our approach will produce much more accurate information about the structure and dynamics of such molecules if we change the parameter $\varepsilon$. Also, the resulting energy eigenvalues are being reduced to the constant mass case energy states in Eq. (25) for the rotation Morse potential in the limit when $\varepsilon \to 0$. The motivation for these generalizations of the potential model with various choices of the parameter $\varepsilon$ provide us a family of energy eigenvalue solutions for the corresponding mass function. Here, in obtaining the bound state energies for $H_2$, $LiH$, $HCl$ and $CO$ molecules, we have provided an alternative method. Our numerical results compare favorably with those obtained using other approximation schemes. The present numerical calculations are obtained by following the Weyl ambiguity ordering $a = 1$ and $\alpha = 0 = \gamma$ or equivalently the Li and Kuhn ambiguity ordering: $a = 0 = \alpha$ and $\gamma = -1/2$. It should be noted that the analytical results presented here allow one to calculate the energy eigenvalues as well as wave functions in a very simple way, with very high accuracy for lower- and higher-excitation vibration levels, enough for most of the applications known until now.

## 5. Acknowledgements

The author highly acknowledges the partial support provided by the Scientific and Technological Research Council of Turkey (TÜBİTAK). He also thanks Professor Shi-Hai Dong for his helpful comments and suggestions.

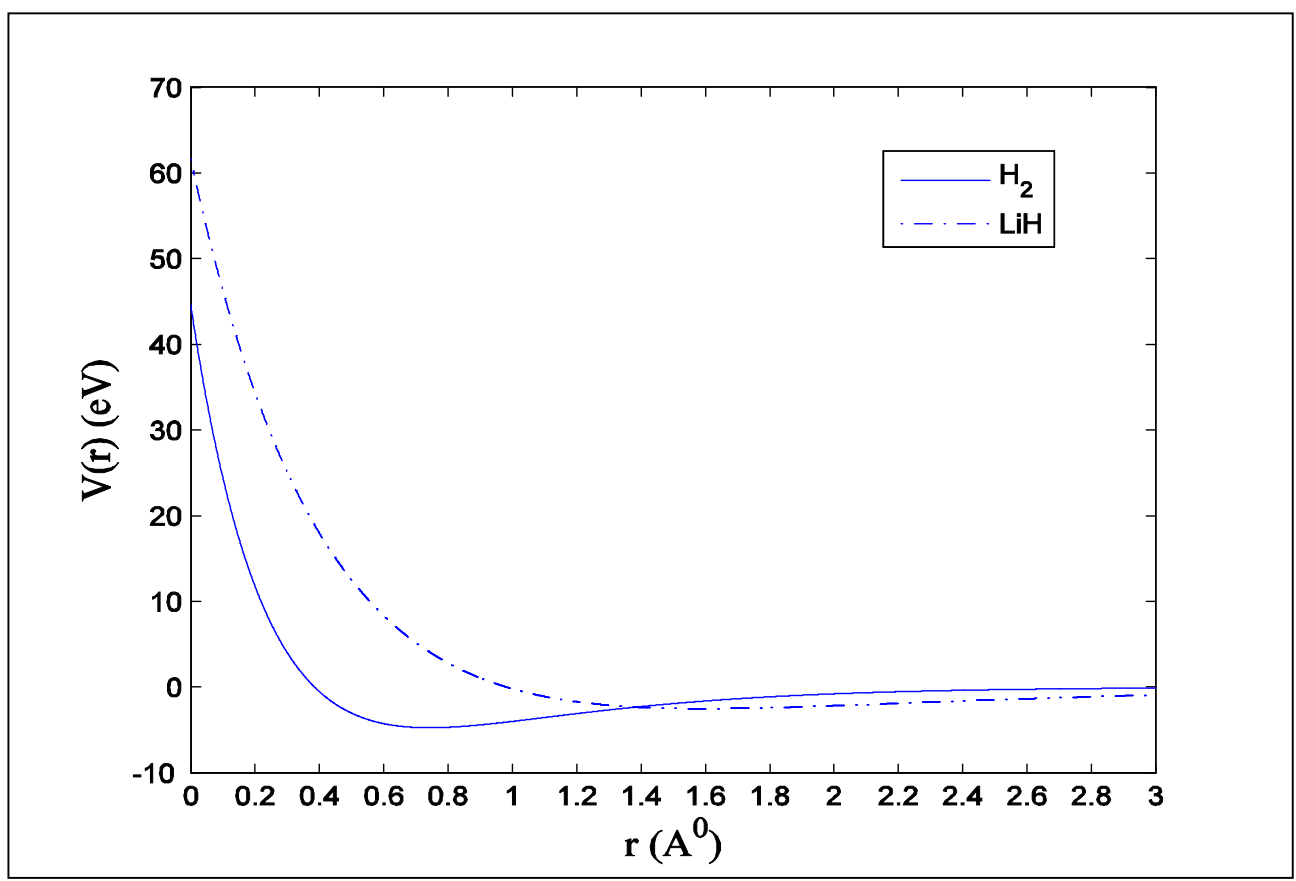

**Figure 1. A plot of Morse potential for $H_2$ and $LiH$ molecules using the set of parameters in Table 1.**

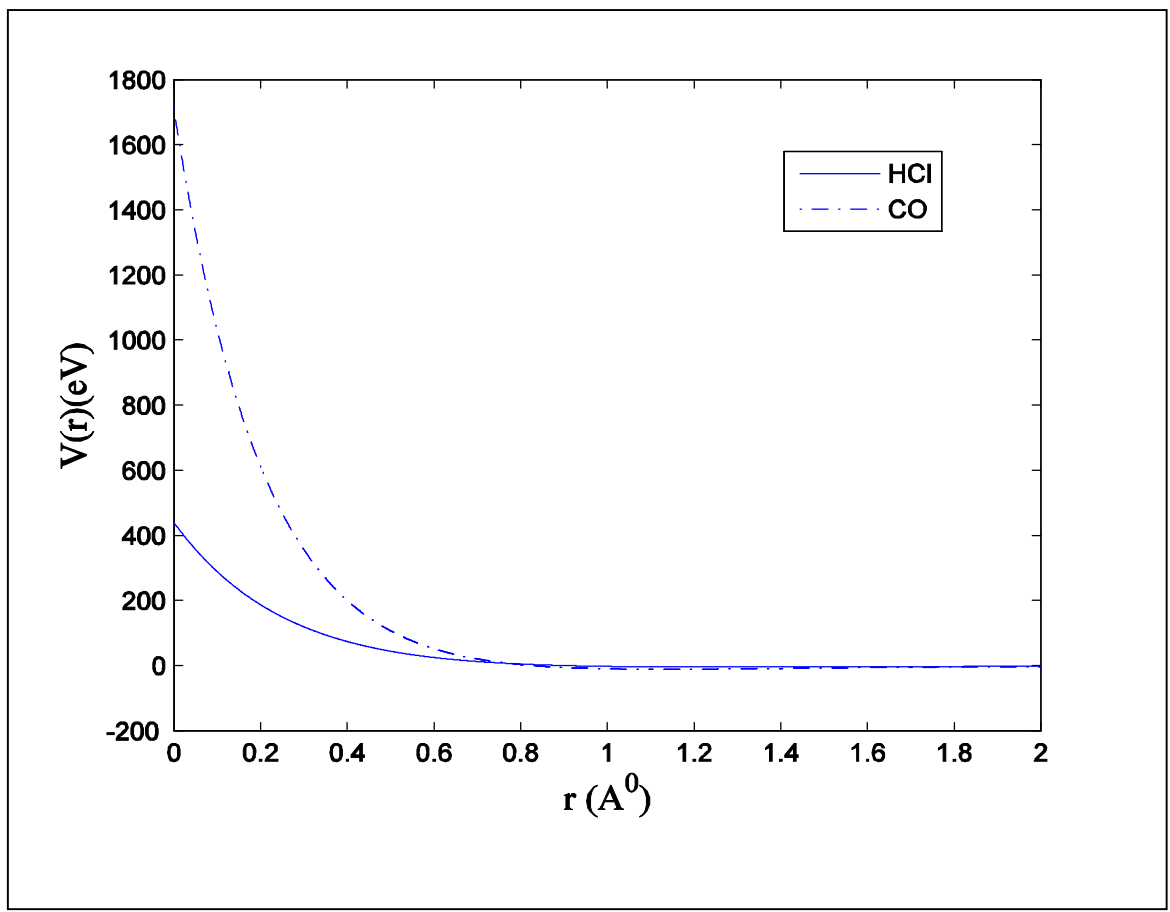

**Figure 2.** A plot of Morse potential for $HCl$ and $CO$ molecules using the set of parameters in Table 1.

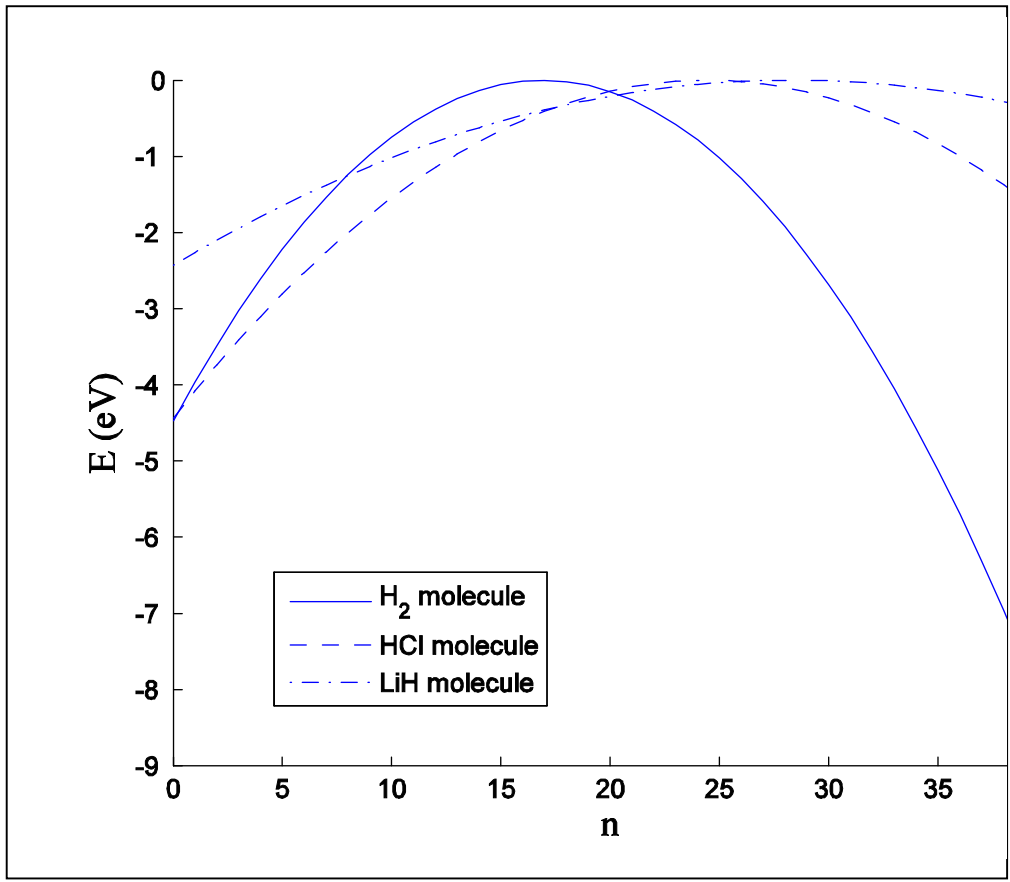

**Figure 3. The s-wave energy spectrum curves as functions of** $n$ **for** $H_2$, *HCl* **and** *LiH* **molecules, respectively.**

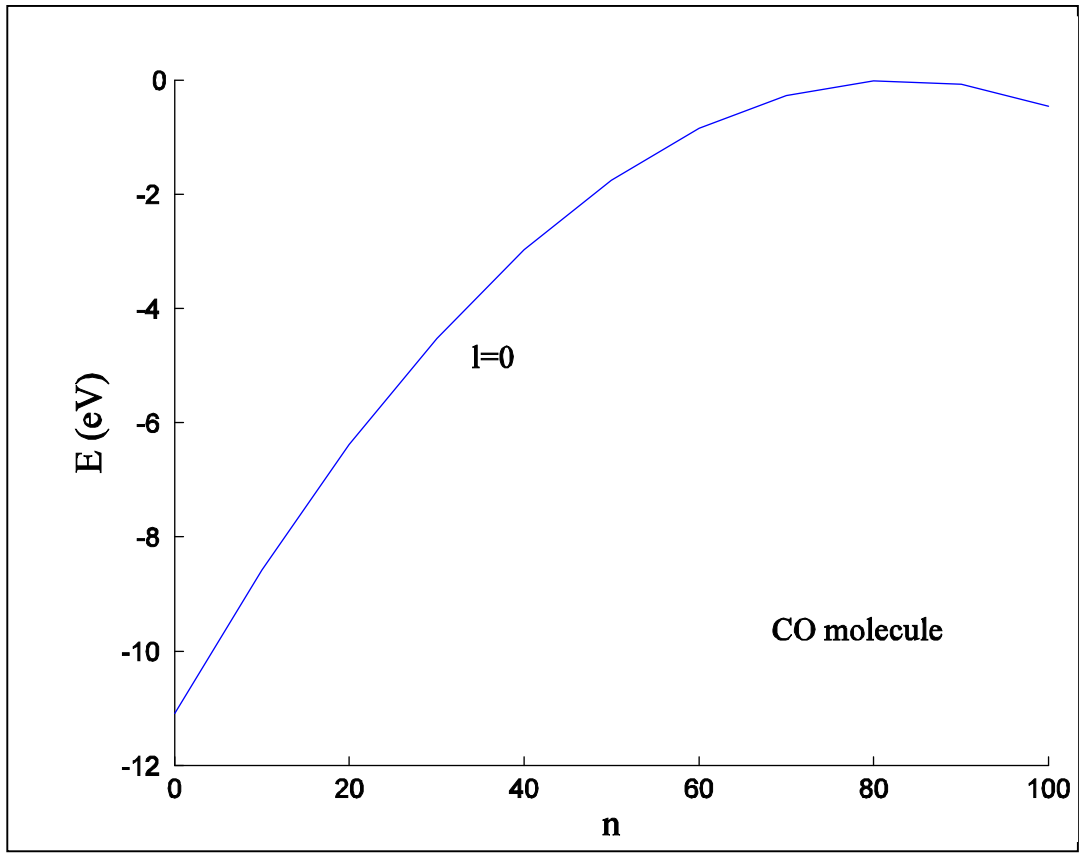

**Figure 4.** The s-wave energy spectrum curves as functions of $n$ for $CO$ molecule.

**Table 1.** Model parameters for the diatomic molecules. $E_0$ is calculated as $E_0 = \hbar^2/(2m_0 r_e^2)$.

| Molecule | Source | $D_e$ (eV) | $r_e$ (A°) | $m$ (amu) | $v = br_e$ | $E_0$ (eV) |
|---|---|---|---|---|---|---|
| $H_2$ | [13] | 4.7446 | 0.7416 | 0.50391 | 1.440558 | $0.754171966 \times 10^{-2}$ |
| $LiH$ | [10] | 2.515287 | 1.5956 | 0.8801221 | 1.7998368 | $0.932764099 \times 10^{-3}$ |
| $HCl$ | [10] | 4.61907 | 1.2746 | 0.9801045 | 2.38057 | $1.312630806 \times 10^{-3}$ |
| $CO$ | [10] | 11.2256 | 1.1283 | 6.8606719 | 2.59441 | $2.393023577 \times 10^{-4}$ |

Table 2: s-state energies of $(-E_n)$ for $H_2$, $LiH$, $HCl$ and $CO$ (in eV) as function of $n$, where $\hbar c = 1973.29\, eV.A^\circ$.

| n | $H_2^a$ | | | $LiH^b$ | | |
|---|---|---|---|---|---|---|
| | This work | TJM [11] | Exact [11] | This work | TJM [11] | Exact [11] |
| 0 | 4.47601 | 4.476013 | 4.476013 | 2.42886 | 2.428863 | 2.428863 |
| 1 | 3.96231 | 3.962315 | 3.962315 | 2.26055 | 2.260548 | 2.260548 |
| 2 | 3.47991 | 3.479919 | 3.479919 | 2.09827 | 2.098276 | 2.098276 |
| 3 | 3.02881 | 3.028824 | 3.028824 | 1.94204 | 1.942047 | 1.942047 |
| 4 | 2.60902 | 2.609030 | 2.609030 | 1.79186 | 1.791862 | 1.791862 |
| 5 | 2.22052 | 2.220537 | 2.220537 | 1.64771 | 1.647720 | 1.647720 |
| 6 | 1.86333 | 1.863345 | 1.863345 | 1.50962 | 1.509621 | 1.509621 |
| 7 | 1.53744 | 1.537455 | 1.537455 | 1.37756 | 1.377565 | 1.377565 |
| 8 | 1.24285 | 1.242866 | 1.242866 | 1.25155 | 1.251552 | 1.251552 |
| 9 | 0.979564 | 0.979579 | 0.979579 | 1.13158 | 1.131583 | 1.131583 |
| 10 | 0.747579 | 0.747592 | 0.747592 | 1.01765 | 1.017656 | 1.017656 |
| 11 | 0.546894 | 0.546907 | 0.546907 | 0.90977 | 0.909773 | 0.909773 |
| 12 | 0.377512 | 0.377523 | 0.377523 | 0.80793 | 0.807934 | 0.807934 |
| 13 | 0.239431 | 0.239441 | 0.239441 | 0.71213 | 0.712137 | 0.712137 |
| 14 | 0.132651 | 0.132659 | 0.132659 | 0.62238 | 0.622384 | 0.622384 |
| 20 | | | | 0.21076 | 0.210771 | 0.210771 |
| 25 | | | | 0.033946 | 0.033949 | 0.033949 |
| $n_{\max}$ | 17 | 17 | 17 | 29 | 29 | 29 |

| n | $HCl^c$ | | | $CO^d$ | | |
|---|---|---|---|---|---|---|
| | This work | TJM [15] | Exact [15] | This work | TJM [15] | Exact [15] |
| 0 | 4.43556 | 4.435564 | 4.435564 | 11.0915 | 11.091535 | 11.091535 |
| 1 | 4.07971 | 4.079710 | 4.079710 | 10.8258 | 10.825822 | 10.825822 |
| 2 | 3.73873 | 3.738734 | 3.738734 | 10.5633 | 10.563330 | 10.563330 |
| 3 | 3.41263 | 3.412635 | 3.412635 | 10.3041 | 10.304060 | 10.304060 |
| 4 | 3.10141 | 3.101414 | 3.101414 | 10.0480 | 10.048011 | 10.048011 |
| 5 | 2.80506 | 2.805071 | 2.805071 | 9.79518 | 9.795184 | 9.795184 |
| 6 | 2.52359 | 2.523605 | 2.523605 | 9.54557 | 9.545578 | 9.545578 |
| 7 | 2.25701 | 2.257018 | 2.257018 | 9.29918 | 9.299193 | 9.299193 |
| 8 | 2.00529 | 2.005307 | 2.005307 | 9.05602 | 9.056030 | 9.056030 |
| 9 | 1.76846 | 1.768475 | 1.768475 | 8.81608 | 8.816089 | 8.816089 |
| 10 | 1.54651 | 1.546520 | 1.546520 | 8.57935 | 8.579369 | 8.579369 |
| 11 | 1.33943 | 1.339442 | 1.339442 | 8.34585 | 8.345870 | 8.345870 |
| 12 | 1.14723 | 1.147243 | 1.147243 | 8.11558 | 8.115593 | 8.115593 |
| 13 | 0.96991 | 0.969921 | 0.969921 | 7.88852 | 7.888538 | 7.888538 |
| 14 | 0.80746 | 0.807476 | 0.807476 | 7.66468 | 7.664704 | 7.664704 |
| 18 | 0.30646 | 0.306476 | 0.306476 | 6.80156 | 6.801582 | 6.801582 |
| 21 | 0.08693 | 0.086940 | 0.086940 | 6.18804 | 6.188066 | 6.188066 |
| 40 | | | | 2.97572 | 2.975752 | 2.975752 |
| 60 | | | | 0.85072 | 0.850621 | 0.850621 |
| $n_{max}$ | 24 | 24 | 25 | 83 | 70 | 83 |

a. The final bound state is $E_{17} = -1.231 \times 10^{-4} eV$.

b. The final bound state is $E_{29} = -1.270 \times 10^{-3} eV$.

c. The final bound state is $E_{24} = -1.303 \times 10^{-3} eV$.

d. The final bound state is $E_{83} = -5.533 \times 10^{-7} eV$.

Table 3: Energy states $(-E_{nl})$ for $H_2$ molecule (in eV) for different values of $l$ and $n$ with a non-constant mass.

| n | l | ε | This work | n | l | ε | This work | n | l | ε | This work |
|---|---|---|---|---|---|---|---|---|---|---|---|
| 0 | 0 | 0.1 | 4.50225 | 5 | 5 | 0.1 | 2.21522 | 6 | 10 | 0.1 | 1.42962 |
|   |   | 0.2 | 4.52872 |   |   | 0.2 | 2.40560 |   |   | 0.2 | 1.60492 |
|   |   | 0.4 | 4.58235 |   |   | 0.4 | 2.85376 |   |   | 0.4 | 2.02927 |
|   |   | 0.6 | 4.63693 |   |   | 0.6 | 3.41631 |   |   | 0.6 | 2.58373 |
|   |   | 0.8 | 4.69249 |   |   | 0.8 | 4.13393 |   |   | 0.8 | 3.32397 |
| 0 | 5 | 0.1 | 4.28381 | 5 | 10 | 0.1 | 1.75241 | 10 | 0 | 0.1 | 0.92597 |
|   |   | 0.2 | 4.30903 |   |   | 0.2 | 1.91639 |   |   | 0.2 | 1.15200 |
|   |   | 0.4 | 4.36013 |   |   | 0.4 | 2.29973 |   |   | 0.4 | 1.81804 |
|   |   | 0.6 | 4.41211 |   |   | 0.6 | 2.77597 |   |   | 0.6 | 2.98370 |
|   |   | 0.8 | 4.46500 |   |   | 0.8 | 3.37626 |   |   | 0.8 | 5.21197 |
| 0 | 10 | 0.1 | 3.74413 | 6 | 0 | 0.1 | 2.05564 | 10 | 5 | 0.1 | 0.77583 |
|   |   | 0.2 | 3.76649 |   |   | 0.2 | 2.27409 |   |   | 0.2 | 0.98431 |
|   |   | 0.4 | 3.81179 |   |   | 0.4 | 2.80923 |   |   | 0.4 | 1.59521 |
|   |   | 0.6 | 3.85783 |   |   | 0.6 | 3.52119 |   |   | 0.6 | 2.65381 |
|   |   | 0.8 | 3.90464 |   |   | 0.8 | 4.49254 |   |   | 0.8 | 4.64747 |
| 5 | 0 | 0.1 | 2.40244 | 6 | 5 | 0.1 | 1.87538 | 10 | 10 | 0.1 | 0.40369 |
|   |   | 0.2 | 2.60453 |   |   | 0.2 | 2.08053 |   |   | 0.2 | 0.57355 |
|   |   | 0.4 | 3.08164 |   |   | 0.4 | 2.58134 |   |   | 0.4 | 1.06486 |
|   |   | 0.6 | 3.68310 |   |   | 0.6 | 3.24410 |   |   | 0.6 | 1.89674 |
|   |   | 0.8 | 4.45420 |   |   | 0.8 | 4.14254 |   |   | 0.8 | 3.41028 |

Table 4: Energy states $(-E_{nl})$ for $LiH$ molecule (in eV) for different values of $l$ and $n$ with a varying mass.

| n | l | ε | This work | n | l | ε | This work | n | l | ε | This work |
|---|---|---|---|---|---|---|---|---|---|---|---|
| 0 | 0 | 0.1 | 2.43768 | 5 | 5 | 0.1 | 1.69554 | 6 | 10 | 0.1 | 1.50275 |
|   |   | 0.2 | 2.44655 |   |   | 0.2 | 1.77179 |   |   | 0.2 | 1.58605 |
|   |   | 0.4 | 2.46442 |   |   | 0.4 | 1.93925 |   |   | 0.4 | 1.77237 |
|   |   | 0.6 | 2.48249 |   |   | 0.6 | 2.12963 |   |   | 0.6 | 1.98976 |
|   |   | 0.8 | 2.50075 |   |   | 0.8 | 2.34720 |   |   | 0.8 | 2.24532 |
| 0 | 5 | 0.1 | 2.41006 | 5 | 10 | 0.1 | 1.63030 | 10 | 0 | 0.1 | 1.11912 |
|   |   | 0.2 | 2.41884 |   |   | 0.2 | 1.70418 |   |   | 0.2 | 1.23396 |
|   |   | 0.4 | 2.43653 |   |   | 0.4 | 1.86636 |   |   | 0.4 | 1.51359 |
|   |   | 0.6 | 2.45441 |   |   | 0.6 | 2.05061 |   |   | 0.6 | 1.88229 |
|   |   | 0.8 | 2.47248 |   |   | 0.8 | 2.26102 |   |   | 0.8 | 2.38007 |
| 0 | 10 | 0.1 | 2.33733 | 6 | 0 | 0.1 | 1.59062 | 10 | 5 | 0.1 | 1.09738 |
|   |   | 0.2 | 2.34587 |   |   | 0.2 | 1.67773 |   |   | 0.2 | 1.21064 |
|   |   | 0.4 | 2.36308 |   |   | 0.4 | 1.87272 |   |   | 0.4 | 1.48633 |
|   |   | 0.6 | 2.38048 |   |   | 0.6 | 2.10049 |   |   | 0.6 | 1.84964 |
|   |   | 0.8 | 2.39806 |   |   | 0.8 | 2.36859 |   |   | 0.8 | 2.33981 |
| 5 | 0 | 0.1 | 1.72034 | 6 | 5 | 0.1 | 1.56642 | 10 | 10 | 0.1 | 1.04023 |
|   |   | 0.2 | 1.79750 |   |   | 0.2 | 1.65247 |   |   | 0.2 | 1.14939 |
|   |   | 0.4 | 1.96699 |   |   | 0.4 | 1.84506 |   |   | 0.4 | 1.41484 |
|   |   | 0.6 | 2.15972 |   |   | 0.6 | 2.06994 |   |   | 0.6 | 1.76417 |
|   |   | 0.8 | 2.38005 |   |   | 0.8 | 2.33456 |   |   | 0.8 | 2.23467 |

Table 5: Energy states $(-E_{nl})$ for $HCl$ molecule (in eV) for different values of $l$ and $n$ with a varying mass case.

| n | l | ε | This work | n | l | ε | This work | n | l | ε | This work |
|---|---|---|---|---|---|---|---|---|---|---|---|
| 0 | 0 | 0.1 | 4.45401 | 5 | 5 | 0.1 | 2.91768 | 6 | 10 | 0.1 | 2.56449 |
|   |   | 0.2 | 4.47257 |   |   | 0.2 | 3.07397 |   |   | 0.2 | 2.73542 |
|   |   | 0.4 | 4.51004 |   |   | 0.4 | 3.42293 |   |   | 0.4 | 3.12585 |
|   |   | 0.6 | 4.54797 |   |   | 0.6 | 3.82903 |   |   | 0.6 | 3.59513 |
|   |   | 0.8 | 4.58636 |   |   | 0.8 | 4.30512 |   |   | 0.8 | 4.16527 |
| 0 | 5 | 0.1 | 4.41514 | 5 | 10 | 0.1 | 2.82723 | 10 | 0 | 0.1 | 1.73795 |
|   |   | 0.2 | 4.43358 |   |   | 0.2 | 2.98020 |   |   | 0.2 | 1.95977 |
|   |   | 0.4 | 4.47079 |   |   | 0.4 | 3.32173 |   |   | 0.4 | 2.52156 |
|   |   | 0.6 | 4.50847 |   |   | 0.6 | 3.71913 |   |   | 0.6 | 3.30651 |
|   |   | 0.8 | 4.54661 |   |   | 0.8 | 4.18495 |   |   | 0.8 | 4.44118 |
| 0 | 10 | 0.1 | 4.31208 | 6 | 0 | 0.1 | 2.68546 | 10 | 5 | 0.1 | 1.70888 |
|   |   | 0.2 | 4.33018 |   |   | 0.2 | 2.86170 |   |   | 0.2 | 1.92854 |
|   |   | 0.4 | 4.36673 |   |   | 0.4 | 3.26429 |   |   | 0.4 | 2.48486 |
|   |   | 0.6 | 4.40373 |   |   | 0.6 | 3.74829 |   |   | 0.6 | 3.26213 |
|   |   | 0.8 | 4.44119 |   |   | 0.8 | 4.33645 |   |   | 0.8 | 4.38560 |
| 5 | 0 | 0.1 | 2.95182 | 6 | 5 | 0.1 | 2.65231 | 10 | 10 | 0.1 | 1.63195 |
|   |   | 0.2 | 3.10936 |   |   | 0.2 | 2.82709 |   |   | 0.2 | 1.84591 |
|   |   | 0.4 | 3.46114 |   |   | 0.4 | 3.22634 |   |   | 0.4 | 2.38780 |
|   |   | 0.6 | 3.87054 |   |   | 0.6 | 3.70629 |   |   | 0.6 | 3.14483 |
|   |   | 0.8 | 4.35052 |   |   | 0.8 | 4.28950 |   |   | 0.8 | 4.23882 |

Table 6: Energy states $(-E_{nl})$ for $CO$ molecule (in eV) for different values of $l$ and $n$ with a varying mass case.

| n | l | ε | This work | n | l | ε | This work | n | l | ε | This work |
|---|---|---|---|---|---|---|---|---|---|---|---|
| 0 | 0 | 0.1 | 11.1049 | 5 | 5 | 0.1 | 9.92329 | 6 | 10 | 0.1 | 9.67726 |
|   |   | 0.2 | 11.1184 |   |   | 0.2 | 10.061  |   |    | 0.2 | 9.83758 |
|   |   | 0.4 | 11.1453 |   |   | 0.4 | 10.3449 |   |    | 0.4 | 10.1701 |
|   |   | 0.6 | 11.1723 |   |   | 0.6 | 10.6408 |   |    | 0.6 | 10.5192 |
|   |   | 0.8 | 11.1994 |   |   | 0.8 | 10.9492 |   |    | 0.8 | 10.8862 |
| 0 | 5 | 0.1 | 11.0978 | 5 | 10 | 0.1 | 9.90487 | 10 | 0 | 0.1 | 8.81530 |
|   |   | 0.2 | 11.1112 |   |    | 0.2 | 10.0424 |    |   | 0.2 | 9.06065 |
|   |   | 0.4 | 11.1381 |   |    | 0.4 | 10.326  |    |   | 0.4 | 9.58165 |
|   |   | 0.6 | 11.1651 |   |    | 0.6 | 10.6214 |    |   | 0.6 | 10.1468 |
|   |   | 0.8 | 11.1922 |   |    | 0.8 | 10.9294 |    |   | 0.8 | 10.7611 |
| 0 | 10 | 0.1 | 11.0787 | 6 | 0 | 0.1 | 9.7024  | 10 | 5 | 0.1 | 8.80864 |
|   |    | 0.2 | 11.0921 |   |   | 0.2 | 9.86303 |    |   | 0.2 | 9.05387 |
|   |    | 0.4 | 11.1190 |   |   | 0.4 | 10.1961 |    |   | 0.4 | 9.57459 |
|   |    | 0.6 | 11.1460 |   |   | 0.6 | 10.5459 |    |   | 0.6 | 10.1394 |
|   |    | 0.8 | 11.1730 |   |   | 0.8 | 10.9135 |    |   | 0.8 | 10.7534 |
| 5 | 0  | 0.1 | 9.9302  | 6 | 5 | 0.1 | 9.69554 | 10 | 10 | 0.1 | 8.79088 |
|   |    | 0.2 | 10.068  |   |   | 0.2 | 9.85609 |    |    | 0.2 | 9.03577 |
|   |    | 0.4 | 10.3521 |   |   | 0.4 | 10.189  |    |    | 0.4 | 9.55577 |
|   |    | 0.6 | 10.648  |   |   | 0.6 | 10.5387 |    |    | 0.6 | 10.1198 |
|   |    | 0.8 | 10.9566 |   |   | 0.8 | 10.9061 |    |    | 0.8 | 10.7329 |